\documentclass[12pt]{article}
\usepackage{amsmath}
\usepackage[colorlinks=true, linkcolor=blue, citecolor=blue, urlcolor=blue]{hyperref}
\usepackage{graphics,graphicx,xcolor}
\def\a{\alpha}
\def\b{\beta}

\def\d{\delta}
\def\e{\epsilon}

\def\g{\gamma}
\def\p{\psi}

\def\r{\rho}
\def\s{\sigma}

\def\he{\hat{e}}

\def\cJ{\mathcal{J}}
\def\be{\begin{equation}}
\def\ee{\end{equation}}
\def\arr{\begin{array}{rll}}
\def\ea{\end{array}}
\def\bea{\begin{eqnarray}}
\def\eea{\end{eqnarray}}
\def\bal{\begin{aligned}}
\def\eal{\end{aligned}}

\def\ic{{\rm i}}
\def\eu{{\rm e}}
\def\N2{$N{=}2$}
\def\diff{{\rm d}}

\def\sfrac#1#2{{\textstyle\frac{#1}{#2}}}
\def\>{\rangle}
\def\<{\langle}
\def\+{\dagger}
\def\={\ =\ }
\def\und{\qquad\textrm{and}\qquad}

\begin{document}

\renewcommand{\thefootnote}{\fnsymbol{footnote}}
\begin{titlepage}
\setcounter{page}{0}
\vskip 1cm
\begin{center}
{\LARGE\bf Spinning extensions of }\\
\vskip 0.5cm
{\LARGE\bf $\boldsymbol{D(2,1;\alpha)}$ superconformal mechanics}\\
\vskip 1.5cm
$
\textrm{\Large Anton Galajinsky\ }^{a}, \quad
\textrm{\Large Olaf Lechtenfeld\ }^{b}
$
\vskip 0.7cm
${}^{a}$ {\it
School of Physics, Tomsk Polytechnic University,
634050 Tomsk, Lenin Ave. 30, Russia} \\
\vskip 0.2cm
${}^{b}$ {\it
Institut f\"ur Theoretische Physik and Riemann Center for Geometry and Physics,\\
Leibniz Universit\"at Hannover, Appelstrasse 2, 30167 Hannover, Germany} \\
\vskip 0.2cm
{Emails: galajin@tpu.ru, lechtenf@itp.uni-hannover.de}
\vskip 0.5cm

\end{center}
\vskip 1cm
\begin{abstract} \noindent
As is known, any realization of SU(2) in the phase space of a dynamical system
can be generalized to accommodate the exceptional supergroup $D(2,1;\alpha)$,
which is the most general $\mathcal{N}{=}\,4$ supersymmetric extension of the
conformal group in one spatial dimension. We construct novel spinning extensions
of $D(2,1;\alpha)$ superconformal mechanics by adjusting the SU(2) generators
associated with the relativistic spinning particle coupled to a spherically symmetric
Einstein--Maxwell background. The angular sector of the full superconformal
system corresponds to the orbital motion of a particle coupled to a symmetric Euler top,
which represents the spin degrees of freedom. This particle moves either on the two-sphere,
optionally in the external field of a Dirac monopole, or in the SU(2) group manifold.
Each case is proven to be superintegrable, and explicit solutions are given.
\end{abstract}

\vspace{0.5cm}

\end{titlepage}

\renewcommand{\thefootnote}{\arabic{footnote}}
\setcounter{footnote}0

\noindent
{\bf 1. Introduction}\\

\noindent
$\mathcal{N}{=}\,4$ superconformal many-body models in one dimension may prove useful in a microscopic description of the near-horizon extremal black holes \cite{GT}. The peculiar features of extended one-dimensional supersymmetry provide another source of inspiration \cite{IKL}. The exceptional supergroup $D(2,1;\alpha)$ plays the key role in this context, because it is the most general $\mathcal{N}{=}\,4$ supersymmetric extension of the conformal group SO(2,1) in one space dimension. The generators of the corresponding Lie superalgebra are associated with time translation, dilatation, special conformal transformation, supersymmetry transformations and their superconformal partners, as well as with two variants of $su(2)$ transformations. One of them is the $R$-symmetry subalgebra, while the other one acts upon fermions only.

In recent works \cite{AG1,AG}, couplings in $\mathcal{N}{=}\,4$ superconformal mechanics have been reconsidered  from the perspective of the $R$-symmetry subgroup. It was argued that
any realization of SU(2) in terms of phase-space functions can be extended to a representation of $D(2,1;\alpha)$. In particular, this allowed one to reproduce the $D(2,1;\alpha)$ supermultplets of type $(3,4,1)$ (two variants), $(4,4,0)$, and $(0,4,4)$ as well as to construct novel couplings (see also \cite{FI}).
The present paper extends the analysis of \cite{AG1,AG} to encompass spin degrees of freedom.

The first attempt to accommodate spin variables within $D(2,1;\alpha)$ superconformal mechanics was made in \cite{FIL} (see also the related earlier work \cite{FIL1}). The $R$-symmetry generators were built in terms of a bosonic SU(2) doublet which parametrizes a two-dimensional sphere. These spin degrees of freedom turned out to be only semi-dynamical, as they are governed by a Wess--Zumino-type action linear in the velocities.  A generalization to the many-body case was proposed in \cite{KL}.

In contrast to the previous studies \cite{FIL,FIL1,KL}, the spinning extensions we build in this work are fully dynamical.
As the principal idea to this end, we borrow the SU(2) generators
of a relativistic spinning particle coupled to a spherically symmetric four-dimensional Einstein--Maxwell background,
which yields a spin sector represented by a symmetric Euler top.
The spin dynamics becomes nontrivially coupled with the orbital motion of the particle.

The work is organized as follows. In the next section, we review the symplectic structure of a spinning particle on a curved background along the lines of \cite{AKH}. In Section~3, spherically symmetric solutions to the four-dimensional Einstein--Maxwell equations are utilized to build three SU(2)-invariant reduced angular Hamiltonian systems in phase space. They describe firstly a particle moving on a two-sphere coupled to a symmetric Euler top, secondly the same system in the external field of a Dirac monopole, and thirdly a particle propagating on the SU(2) group manifold and interacting with a symmetric Euler top. We emphasize that the underlying Poisson-structure relations are not canonical and involve an arbitrary real parameter $a\in (0,1)$, which is linked to the $g_{00}$ component of the original background metric. In the reduced SU(2) mechanics it determines the moments of inertia of the symmetric Euler top. Each case is shown to be integrable, and the corresponding solutions to the equations of motion are displayed in Section~4.
The rotation to a reference frame better adapted to the orbital motion is discussed in Section~5, including some subtleties.
Section~6 uses our spin-orbit $su(2)$ generators to build novel spinning extensions of $D(2,1;\alpha)$ superconformal mechanics along the lines in \cite{AG}. In the concluding Section~7 we summarize our results and discuss possible further developments.

Throughout the paper a summation over repeated indices is understood. We use units in which $c=1$ and $G=1$. In Section~2 the Greek letters refer to four-dimensional curved spacetime indices, while in Section~6 they designate SU(2) doublet representations. The relation between spherical and Cartesian coordinates and our SU(2) spinor conventions are gathered in an Appendix.

\vspace{0.5cm}

\noindent
{\bf 2. Symplectic structure of a spinning particle on a curved background}\\

\noindent
The phase space of a spinning particle on a curved background is parametrized by the canonical pair $(x^\mu,p_\mu)$ and self-conjugate spin variables $S^{\mu\nu}=-S^{\nu\mu}$ with $\mu,\nu=0,1,2,3$. It is endowed with
the symplectic structure \cite{AKH}
\be \label{Br}
\bal
&
\{x^\mu,p_\nu \}={\delta^\mu}_\nu, \quad \{p_\mu,p_\nu\}=-\sfrac 12 R_{\mu\nu\lambda\sigma} S^{\lambda\sigma}, \quad \{S^{\mu\nu},p_\lambda \}=\Gamma^\mu_{\lambda\sigma} S^{\nu\sigma}-\Gamma^\nu_{\lambda\sigma} S^{\mu\sigma},
\\[2pt]
&
\{S^{\mu\nu},S^{\lambda\sigma} \}=g^{\mu\lambda} S^{\nu\sigma}+g^{\nu\sigma} S^{\mu\lambda}-g^{\mu\sigma} S^{\nu\lambda}-g^{\nu\lambda} S^{\mu\sigma},
\eal
\ee
where $g^{\mu\nu}$ is the inverse metric tensor, $\Gamma^\mu_{\lambda\sigma}$ are the Christoffel symbols, and
$R_{\mu\nu\lambda\sigma}$ is the Riemann tensor.\footnote{
Our conventions are ${R^{\alpha}}_{\beta\gamma\delta}=\partial_\gamma \Gamma^\alpha_{\delta\beta}-\partial_\delta \Gamma^\alpha_{\gamma\beta}+\Gamma^\alpha_{\gamma\sigma}\Gamma^\sigma_{\delta\beta}-\Gamma^\alpha_{\delta\sigma}\Gamma^\sigma_{\gamma\beta}$ and $\Gamma^\alpha_{\beta\gamma}=\frac 12 g^{\alpha\lambda} (\partial_\beta g_{\lambda\gamma}+
\partial_\gamma g_{\lambda\beta}-\partial_\lambda g_{\beta\gamma} )$.}
The Jacobi identities are fulfilled as a consequence of the Bianchi identity
$\nabla_\alpha R_{\lambda\sigma \beta\gamma}+\nabla_\beta R_{\lambda\sigma \gamma\alpha}+\nabla_\gamma R_{\lambda\sigma \alpha\beta}=0$
and the fact that the metric is covariantly constant.

In what follows we will need the following statement. Let $\xi_1^\mu \partial_\mu$, $\xi_2^\mu \partial_\mu$ and $\xi_3^\mu \partial_\mu$ be three Killing vector fields obeying
\be
[\xi_1^\lambda \partial_\lambda,\xi_2^\nu \partial_\nu]=\xi_3^\mu \partial_\mu \qquad\textrm{with}\qquad
\xi_3^\mu=\xi_1^\sigma \partial_\sigma \xi_2^\mu-\xi_2^\sigma \partial_\sigma \xi_1^\mu.
\ee
Then the phase-space functions
\be\label{Int}
\mathcal{J}(\xi)\=\xi^\mu p_\mu+\sfrac12 \nabla_\mu \xi_\nu S^{\mu\nu}
\ee
satisfy a similar relation under the bracket (\ref{Br}), namely
\be
\{\mathcal{J}(\xi_1),\mathcal{J}(\xi_2) \}\=-\mathcal{J}(\xi_3).
\ee
The proof is straightforward and relies upon the relation
\be
\nabla_\lambda \nabla_\mu \xi_\nu S^{\mu\nu}\={R^\gamma}_{\lambda\mu\nu}\,\xi_\gamma S^{\mu\nu},
\ee
which is valid for an arbitrary Killing vector $\xi_\nu$ as a consequence of $\nabla_\mu\xi_\nu=-\nabla_\nu\xi_\mu$, $[\nabla_\mu,\nabla_\nu]\xi_\gamma=-{R^\lambda}_{\gamma\mu\nu} \xi_\lambda$, and ${R^\gamma}_{\lambda\mu\nu}+{R^\gamma}_{\mu\nu\lambda}+{R^\gamma}_{\nu\lambda\mu}=0$.
Hence, a background invariance under some Lie group implies a natural action of the same group in the phase space endowed with the symplectic structure (\ref{Br})
(see also the discussion in \cite{AKH}).

\newpage

\noindent
{\bf 3. Spherically symmetric backgrounds and reduced SU(2) mechanics}\\

\noindent
The unique spherically symmetric solution of the four-dimensional vacuum Einstein equations is the Schwarzschild black hole metric
\be\label{Sch}
\diff s^2\=\left(1-\sfrac{2M}{r} \right) \diff t^2-{\left(1-\sfrac{2M}{r} \right)}^{-1} \diff r^2-r^2 (\diff \theta^2+\sin^2\!\theta\,\diff \phi^2),
\ee
where $M$ is the mass and $r>2M$. 
Its spatial Killing vector fields generating an $su(2)$ algebra read
\be\label{KV}
-\sin{\phi}\,\partial_\theta-\cot{\theta} \cos{\phi}\,\partial_\phi, \qquad
\cos{\phi}\,\partial_\theta-\cot{\theta} \sin{\phi}\,\partial_\phi, \qquad \partial_\phi.
\ee

Computing the geometric characteristics $\Gamma^\mu_{\lambda\sigma}$, $R_{\mu\nu\lambda\sigma}$ and evaluating the phase-space functions (\ref{Int}), one finds that $p_t$, $p_r$, $S^{tr}$, $S^{t\theta}$, and $S^{t \phi}$ do not contribute to (\ref{Int}).
It is therefore consistent to reduce the spinning-particle dynamics on this background to a spherical one
by ignoring the coordinate time~$t$ and regarding the radial variable~$r$ as a fixed external parameter.
The ensuing reduced SU(2) mechanical system is then governed only by the two angular variable-momentum pairs $(\theta,p_{\theta})$ and $(\phi,p_\phi)$ as well as
spin vector $\vec{J}$ with components $(J_r,J_\theta,J_\phi)$ built from the triple $(S^{r\theta},S^{r\phi},S^{\theta\phi})$.
Abbreviating
\be
1-\sfrac{2M}{r}=a^2 \qquad\textrm{with}\quad a\in (0,1)
\ee
and denoting
\be\label{redif}
J_\phi=\sfrac{r}{a}S^{r\theta}=:J_1, \qquad J_r=r^2 \sin{\theta}\,S^{\theta\phi}=:J_2, \qquad J_\theta=\sfrac{r}{a}\sin{\theta}\,S^{r\phi}=:J_3,
\ee
one reduces (\ref{Br}) to the Poisson structure
\begin{align}\label{sr}
&
\{\theta,p_\theta\}=1, && \{\phi,p_\phi\}=1, && \{p_\theta,p_\phi\}=(1-a^2)J_2  \sin{\theta},
\nonumber\\[4pt]
&
\{J_1,p_\phi\}=J_3 \cos{\theta}-a J_2 \sin{\theta}, && \{J_2,p_\theta\}=-a J_3, && \{J_2,p_\phi\}=a J_1 \sin{\theta},
\\[4pt]
&
\{J_3,p_\theta\}=a J_2, && \{J_3,p_\phi\}=-J_1 \cos{\theta}, && \{J_i,J_j\}=\epsilon_{ijk}J_k,
\nonumber
\end{align}
where $\epsilon_{ijk}$ is the Levi--Civita symbol with $\epsilon_{123}=1$. It is straightforward to verify that the Jacobi identities are satisfied for (\ref{sr}).

The $su(2)$ generators constructed from~(\ref{Int}) now acquire the form
\bea\label{charges}
&&
\mathcal{J}_1\=-(p_\theta-a J_1) \sin{\phi}-\left(\frac{p_\phi}{ \sin{\theta}}-a J_3\right) \cos{\theta}\cos{\phi}-J_2 \sin{\theta}\cos{\phi},
\nonumber\\[4pt]
&&
\mathcal{J}_2\=(p_\theta-a J_1)  \cos{\phi}-\left(\frac{p_\phi}{ \sin{\theta}}-a J_3\right) \cos{\theta}\sin{\phi}-J_2 \sin{\theta}\sin{\phi},
\\[4pt]
&&
\mathcal{J}_3\=\left(\frac{p_\phi}{ \sin{\theta}}-a J_3\right)\sin{\theta}-J_2 \cos{\theta}.
\nonumber
\eea
Decomposing into orbital and spin angular momentum, we may write
\be
\vec{\cJ}=\vec{L}+\vec{J} \qquad\textrm{with}\qquad
\vec{L}\= p_\theta\,\he_\phi - \frac{p_\phi}{\sin\theta}\he_\theta \und
\vec{J}\= -a J_1\,\he_\phi + a J_3\,\he_\theta - J_2\,\he_r,
\ee
where $(\he_r,\he_\theta,\he_\phi)$ are the standard local orthonormal basis vectors
associated to three-dimen\-sional spherical coordinates and given in the Appendix.

To define our reduced dynamical system we need to specify a Hamiltonian which generates a proper-time evolution.
A minimal and natural choice is the $su(2)$ Casimir element
\be\label{H1}
H\=\sfrac12\mathcal{J}_i\mathcal{J}_i
\=\sfrac12\bigl(\vec{L}+\vec{J})^2
\=\sfrac12 \Bigl(\bigl(p_\theta-a J_1 \bigr)^2+\bigl( \frac{p_\phi}{\sin{\theta}}-a J_3 \bigr)^2+J_2^2 \Bigr).
\ee
Two limiting cases are worth mentioning.
On the one hand, in the absence of the spin degrees of freedom $\vec J$ this Hamiltonian describes a free particle of unit mass moving on a two-dimensional unit sphere.
On the other hand, discarding the angular canonical pairs $(\theta,p_{\theta})$ and $(\phi,p_\phi)$,
one reveals a symmetric free Euler top:
\be
\dot{J}_i = \{J_i,H\} \qquad\Rightarrow\qquad
{\dot J}_1=(1{-}a^2)J_2 J_3, \qquad {\dot J}_2=0, \qquad {\dot J}_3=-(1{-}a^2)J_1 J_2.
\ee
Hence, (\ref{sr}) and (\ref{H1}) couples these two systems and describes a spinning particle on a two-sphere.
The composite system is superintegrable.
By construction, the total angular momentum vector $\vec{\cal J}$ is conserved.
Four functionally independent integrals of motion in involution included
$(\vec\cJ^2{=}2H,\cJ_3,J_1^2{+}J_3^2,J_2)$.
A fifth functionally independent conserved quantity, $\vec{L}^2$, is in involution with $J_iJ_i$
but not with $\vec{J}^2=a^2(J_1^2{+}J_3^2){+}J_2^2$. Hence, the system in superintegrable.
In Minkowski space, $a{=}1$, one may alternatively choose the Liouville set $(\vec\cJ^2{=}2H,\cJ_3,\vec{L}^2,\vec{J}^2)$.
This list is in complete analogy with the well-known spin-orbit coupling problem in quantum mechanics.
The radial deformation of the metric, $|g_{rr}|=a^{-2}$, modifies this picture.

Turning to the four-dimensional Einstein--Maxwell equations, the spherically symmetric solution is given by the Reissner--Nordstr\"om black hole
\be\label{RN}
\diff s^2\=(1-\sfrac{2M}{r}+\sfrac{Q^2}{r^2})\,\diff t^2- (1-\sfrac{2M}{r}+\sfrac{Q^2}{r^2} )^{-1}\diff r^2-r^2 (\diff\theta^2+\sin^2\!{\theta}\,\diff \phi^2), \quad
A\=\sfrac{Q}{r}\diff t,
\ee
where $M$ is the mass and $Q$ the electric charge. One can repeat the analysis above and reproduce the same relations (\ref{sr}) and (\ref{charges})
with the obvious modification of the external parameter,
\be
a^2\=1-\sfrac{2M}{r}+\sfrac{Q^2}{r^2}.
\ee
However, if the black hole also carries a magnetic charge $q$, the latter contributes to the $su(2)$ generators,
\bea\label{charges1}
&&
\mathcal{J}_1\=-(p_\theta-a J_1) \sin{\phi}-\left(\frac{p_\phi}{ \sin{\theta}}-a J_3\right) \cos{\theta}\cos{\phi}-J_2 \sin{\theta}\cos{\phi}+q\,\frac{\cos{\phi}}{\sin{\theta}},
\nonumber\\[4pt]
&&
\mathcal{J}_2\=(p_\theta-a J_1)  \cos{\phi}-\left(\frac{p_\phi}{ \sin{\theta}}-a J_3\right) \cos{\theta}\sin{\phi}-J_2 \sin{\theta}\sin{\phi}+q\,\frac{\sin{\phi}}{\sin{\theta}},
\\[4pt]
&&
\mathcal{J}_3\=\left(\frac{p_\phi}{ \sin{\theta}}-a J_3\right)\sin{\theta}-J_2 \cos{\theta}.
\nonumber
\eea
The structure relations (\ref{sr}) remain intact. The associated $su(2)$ mechanics is governed by the Hamiltonian
\be\label{H2}
H\=\sfrac12\mathcal{J}_i\mathcal{J}_i\=\sfrac 12 \Bigl(\bigl(p_\theta-a J_1 \bigr)^2+\bigl( \frac{p_\phi}{\sin{\theta}}-q \cot{\theta}-a J_3 \bigr)^2+(J_2-q)^2 \Bigr),
\ee
which describes a spinning particle moving on a unit two-sphere in the external field of a Dirac monopole.

As $q$ in (\ref{charges1}) and (\ref{H2}) is a constant, one can build one more realization of $su(2)$. For this we introduce an extra canonical pair $(\xi,p_\xi)$,
extend the structure relations (\ref{sr}) by
\be
\{\xi,p_\xi\}=1
\ee
and implement an oxidation with respect to $q$ by replacing
\be
q \ \rightarrow\ p_\xi.
\ee
The resulting Hamiltonian
\be\label{H3}
H\=\sfrac12\mathcal{J}_i\mathcal{J}_i\=\sfrac 12 \Bigl(\bigl(p_\theta-a J_1 \bigr)^2+\bigl( \frac{p_\phi}{\sin{\theta}}-p_\xi\cot{\theta}-a J_3 \bigr)^2+(J_2-p_\xi)^2 \Bigr)
\ee
describes a spinning particle propagating on the group manifold of SU(2). It is straightforward to verify that the corresponding $su(2)$ generators
\bea\label{charges2}
&&
\mathcal{J}_1\=-(p_\theta-a J_1) \sin{\phi}-\left(\frac{p_\phi}{ \sin{\theta}}-a J_3\right) \cos{\theta}\cos{\phi}-J_2 \sin{\theta}\cos{\phi}+ \frac{p_\xi \cos{\phi}}{\sin{\theta}},
\nonumber\\[4pt]
&&
\mathcal{J}_2\=(p_\theta-a J_1)  \cos{\phi}-\left(\frac{p_\phi}{ \sin{\theta}}-a J_3\right) \cos{\theta}\sin{\phi}-J_2 \sin{\theta}\sin{\phi}+\frac{p_\xi \sin{\phi}}{\sin{\theta}},
\\[4pt]
&&
\mathcal{J}_3\=\left(\frac{p_\phi}{ \sin{\theta}}-a J_3\right)\sin{\theta}-J_2 \cos{\theta}
\nonumber
\eea
reduce to the vector fields dual to the conventional left-invariant one-forms defined on the group manifold in case the spin degrees of freedom are absent.
Like its reduction (\ref{H2}), the extended model is superintegrable.
Five functionally independent integrals of motion in involution are given by
$(\vec{\cJ}^2{=}2H,\mathcal{J}_3,p_\xi,J_1^2{+}J_3^2,J_2)$,
an additional integral is still $\vec{L}^2$.

Concluding this section, we note that the Schwarzschild profile $a^2=1-\sfrac{2M}{r}$ of our spherically symmetric background appears to be irrelevant
for obtaining the Poisson structure (\ref{sr}) or the $su(2)$ realization (\ref{charges}).
Indeed, (\ref{Sch}) or (\ref{RN}) may be generalized to a generic static and spherically symmetric metric
\be\label{metric}
\diff s^2\=f(r)\,\diff t^2-f(r)^{-1}\diff r^2-r^2 (\diff \theta^2+\sin^2\!{\theta}\,\diff \phi^2)
\qquad\textrm{with $f(r)>0$ arbitrary}.
\ee
Repeating the analysis above, one arrives at the same expressions (\ref{redif})--(\ref{charges}) with the obvious substitution $a^2 \to f$.

If desirable, (\ref{metric}) can be incorporated within a general relativistic framework as the so called regular black-hole solution.
It suffices to consider Einstein gravity coupled to a variant of nonlinear electrodynamics,
\be
S\=-\sfrac{1}{16 \pi}\int\!\!\diff^4 x\,\sqrt{-g}\,\bigl(R+\mathcal L(F^2)\bigr)
\ee
where $R$ denotes the Riemann curvature scalar and $F^2=F_{\mu \nu} F^{\mu \nu}$ with the gauge field-strength tensor $F_{\mu \nu}$,
and then fix the form of the function $\mathcal L$ from the Einstein--Maxwell equations (for more details see \cite{FW} and references therein).

\vspace{0.5cm}

\newpage

\noindent
{\bf 4. Dynamics of the SU(2) mechanics}\\

\noindent
The reduced SU(2) mechanics is integrable and hence can be solved by quadrature.
Let us start with the model (\ref{H1}) and denote the canonical time variable by~$t$.
Taking into account the Poisson-structure relations and the Hamiltonian, one obtains the equations of motion
\be\label{sup}
\bal
&
\dot\theta\=p_\theta-a J_1,
&&
{\dot J}_1\=J_3\,\bigl(J_2+\dot\phi \cos{\theta} \bigr),
\\[4pt]
&
\dot\phi\=\frac{1}{\sin{\theta}}\Bigl(\frac{p_\phi}{ \sin{\theta}}-a J_3 \Bigr),  \qquad\qquad
&&
{\dot J}_3\=-J_1\,\bigl(J_2+\dot\phi\cos{\theta} \bigr),
\eal
\ee
while (\ref{charges}) and (\ref{H1}) allow one to express $p_\phi$ and $p_\theta$ in terms of the other variables and conserved quantities,
\be\label{sup1}
\frac{p_\phi}{ \sin{\theta}}-a J_3\=\frac{\mathcal{J}_3}{\sin{\theta}}+J_2 \cot{\theta} \ \quad\textrm{and}\quad\
p_\theta-a J_1\=\pm\sqrt{2H - J_2^2-\Bigl(\frac{\mathcal{J}_3}{\sin{\theta}}+J_2 \cot{\theta} \Bigr)^2}.
\ee
Taking into account the rightmost equation in (\ref{sup1}), the differential equation for $\theta$ in (\ref{sup}) can be readily integrated to yield
\be\label{T}
\cos{\theta}(t)\=\sqrt{\Bigl(1-\frac{J_2^2}{2H} \Bigr)\Bigl(1-\frac{\mathcal{J}_3^2}{2H} \Bigr)}\,\cos{\bigl(\sqrt{2H} (t{-}t_0)\bigr)}
-\frac{\mathcal{J}_3 J_2}{2H},
\ee
where $t_0$ is a constant of integration. The time evolution of $\phi$ is tied to that of $\theta$,
\be\label{Phi}
\phi(t)\=\phi_0+\int_{t_0}^t\!\diff \tau\;\frac{\mathcal{J}_3+J_2 \cos{\theta(\tau)}}{\sin^2\!{\theta(\tau)}},
\ee
$\phi_0$ being another constant of integration. Note that $J_2^2\leq 2H$ and $\mathcal{J}_3^2\leq 2H$ as a consequence of (\ref{H1}).

The orbital behaviour of the system becomes more transparent
by observing that
\be\label{cone}
\he_r\cdot\vec\cJ\=x_i\,{\cal J}_i \= -J_2 \= \textrm{constant},
\ee
with $\he_r$ and $x_i$ given in the Appendix.
Since $\he_r$ points to the location $(\theta,\phi)$ of the particle, the latter traces a circular orbit,
which is given by the intersection of our unit two-sphere with a cone
whose axis is determined by the conserved angular momentum vector $\vec{\mathcal{J}}$.
The apex semi-angle $\alpha$ depends on the conserved component $J_2$ of the spin vector $\vec{J}$
and the energy $\sqrt{2H}=|\vec{\mathcal{J}}|$,
\be\label{angle}
\cos{\alpha}=-\frac{J_2}{\sqrt{2H}}.
\ee
If $J_2=0$ the cone opens to the plane $x_i \mathcal{J}_i=0$, and the orbit becomes a great circle. The orbital circular motion is uniform, as follows from
\be \label{uniformorbit}
\dot{x}_i\dot{x}_i \= \dot{\theta}^2+\sin^2\!\theta\,\dot{\phi}^2 \= 2H - J_2^2 \= \textrm{constant}.
\ee

Turning to the spin sector, the integral of motion $J_1^2+J_3^2 =: R^2$
implies that $J_1$ and $J_3$ can be represented in the form~\footnote{
Alternatively, the time reparametrization 
$t \to T=\frac{1}{J_2}\int_0^t d\tau \bigl(J_2+\dot\phi(\tau) \cos{\theta(\tau)} \bigr)$ 
brings the right column in (\ref{sup}) to the standard Euler form. 
The motion of the spin vector is uniform and only with respect to the redefined temporal variable: 
$(\partial_T J_1)^2+(\partial_T J_2)^2+(\partial_T J_3)^2=(J_1^2{+}J_3^2)\,J_2^2=\textrm{constant}$. }
\be\label{J1J2}
J_1(t)=R \cos{\Omega(t)} \und J_3(t)=R\sin{\Omega(t)}.
\ee
Substituting these expressions into the right column of (\ref{sup}), one links $\Omega$ to $\theta$,
\be\label{Om}
\Omega(t)\=\Omega_0-\int_{t_0}^t\!\diff\tau\;\frac{J_2+\mathcal{J}_3\cos{\theta(\tau)}}{\sin^2\!{\theta(\tau)}},
\ee
where $\Omega_0$ is a constant of integration. The spin vector $\vec{J}$
precesses around the radial direction, which corresponds to the 2-direction in the spin subspace
parametrized by $(J_1,J_2,J_3)$. Remarkably enough, the angular precession velocity $\dot\Omega$ is tied to the orbital motion of the particle on the sphere.
As an illustration, below we display graphs of the angular velocities $\dot\theta(t)$, $\dot\phi(t)$ and $\dot\Omega(t)$
for a particular solution.
\begin{figure}[ht]
\begin{center}
\resizebox{0.6\textwidth}{!}{%
\includegraphics{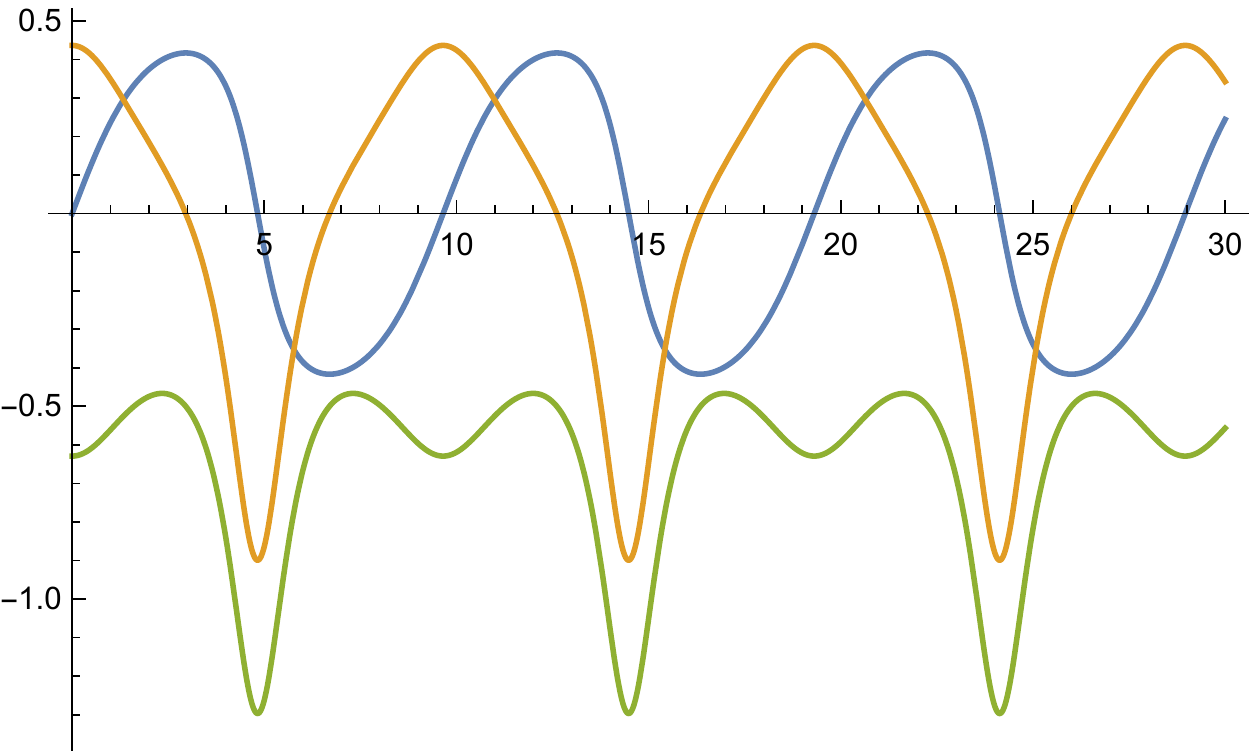}}\vskip-4mm
\caption{\small  The angular velocities
$\dot\theta(t)$, $\dot\phi(t)$ and $\dot\Omega(t)$ for a particular solution, characterized by
$(\cJ_1,\cJ_2,\cJ_3)=(\frac 12,\frac 13,\frac 14)$ and $J_2=\frac 12$.
The blue curve shows $\dot\theta$, the yellow one $\dot\phi$, the green one $\dot\Omega$. }
\label{fig1}
\end{center}
\end{figure}

A particularly simple solution arises if one chooses the initial conditions such that
\be \label{special}
\mathcal{J}_1=\mathcal{J}_2=0.
\ee
In this case we learn that
\be
p_\theta-a J_1 \= 0 \und \Bigl(\frac{p_\phi}{\sin\theta}-a J_3\Bigr) + J_2\tan\theta \= 0,
\ee
which immediately implies that
\be
\dot\theta=0 \und \dot\phi = -\frac{J_2}{\cos\theta} = {\cJ}_3 = \sqrt{2H} = \textrm{constant}
\qquad\Rightarrow\qquad \dot\Omega = 0.
\ee
Hence, in this special case the spin vector $\vec{J}$ is constant (the spin frame is just carried around with the particle),
and so are $p_\phi$ and $p_\theta$. The orbital and spin motion are decoupled.
One might think that the initial condition (\ref{special}) can always be achieved by choosing adapted coordinates via
rotating to a reference frame where $\vec\cJ$ points in the 3-direction.
However, this is not so, as we shall argue in the following section.

A similar analysis can be carried out for the model (\ref{H2}). It is straightforward to verify that (\ref{T})--(\ref{Om}) maintain their form, provided $J_2$ on the right-hand side is replaced by $J_2-q$. In particular, a particle on $S^2$ moves along a great circle if $J_2=q$.

Finally, because (\ref{H3}) is an oxidation of (\ref{H2}) and $p_\xi$ is a constant of the motion, solutions to the equations of motion for $(\theta,\phi,J_1,J_3)$ read as in (\ref{T}), (\ref{Phi}), (\ref{J1J2}) and (\ref{Om}) with the obvious substitution
\be
J_2 \quad \rightarrow \quad J_2-p_\xi
\ee
on the right-hand side of all formul\ae. The equation of motion for the extra angular variable $\xi$ has the general solution
\be
\xi(t)=\xi_0+\Omega(t),
\ee
where $\xi_0$ is a constant of integration, and $\Omega$ reads as in (\ref{Om}) with $J_2$ replaced by $J_2-p_\xi$.

Concluding this section, we note that for a general solution the dimensionless parameter $a$ enters only the relations linking the momenta $(p_\theta,p_\phi)$ to the velocities $(\dot\theta,\dot\phi)$ and the spin degrees of freedom $(J_1,J_3)$ (see the left column in (\ref{sup})). Although $a$ is involved in the formal Hamiltonian formulation, it does not influence the qualitative behaviour of the spinning particle moving on $S^2$ or SU(2).
\vspace{0.5cm}

\noindent
{\bf 5. Rotating the reference frame}\\

\noindent
When solving the equations of motion of a free particle on $S^2$, it is customary to exploit the SU(2) invariance for passing to the reference frame in which the conserved angular momentum vector $\vec{\mathcal{J}}'$ is directed along the $x'_3$--axis.
The condition $\he'_r\cdot\vec{\mathcal{J}}'=0$ then implies that the particle moves along the equator. Alternatively, one can substitute $\theta'=\frac{\pi}{2}$ directly into the Lagrangian and reveal the uniform circular motion $\phi'(t)=\phi_0+p'_\phi t$, where $p'_\phi=\mathcal{J}'_3$ and $\phi'_0$ are constants of integration.

The systems described in the preceding section are more complex. For one thing, they are intrinsically Hamiltonian, and one cannot just substitute $\theta'=\mbox{const}$ into a Lagrangian. For another, the infinitesimal form
\be\label{inf}
\delta _\epsilon A=\{A,\mathcal{J}_i\}\,\epsilon_i
\ee
for SU(2) transformations of an arbitrary phase-space function $A$, where $\epsilon_i$ is an infinitesimal parameter
and $\mathcal{J}_i$ is taken from (\ref{charges}), (\ref{charges1}) or (\ref{charges2}), affects also the spin degrees of freedom $J_i$.
Hence, the rotation takes place in the {\it full\/} phase space.

Although we do not have at hand the explicit canonical transformation generated by a finite analog of (\ref{inf}),
it is clear what to start with.
For definiteness let us focus on the model (\ref{charges}) and (\ref{H1}) and assume $\mathcal{J}^2_1+\mathcal{J}^2_2 \ne 0$.
One can introduce the conserved rotation matrices
\be
R_2= \left(
                                \begin{array}{cccc}
                                1 & 0 & 0 \\
                                0 &   \frac{\mathcal{J}_3}{\sqrt{\mathcal{J}_1^2+\mathcal{J}_2^2+\mathcal{J}_3^2}} & \frac{\sqrt{\mathcal{J}_1^2+\mathcal{J}_2^2}}{\sqrt{\mathcal{J}_1^2+\mathcal{J}_2^2+\mathcal{J}_3^2}} \\
                               0 &   -\frac{\sqrt{\mathcal{J}_1^2+\mathcal{J}_2^2}}{\sqrt{\mathcal{J}_1^2+\mathcal{J}_2^2+\mathcal{J}_3^2}} & \frac{\mathcal{J}_3}{\sqrt{\mathcal{J}_1^2+\mathcal{J}_2^2+\mathcal{J}_3^2}}  \\
                                \end{array}
                              \right)
\und
 R_1= \left(
                                \begin{array}{cccc}
                                  -\frac{\mathcal{J}_2}{\sqrt{\mathcal{J}_1^2+\mathcal{J}_2^2}} & \frac{\mathcal{J}_1}{\sqrt{\mathcal{J}_1^2+\mathcal{J}_2^2}}  & 0  \\
                                  -\frac{\mathcal{J}_1}{\sqrt{\mathcal{J}_1^2+\mathcal{J}_2^2}} & -\frac{\mathcal{J}_2}{\sqrt{\mathcal{J}_1^2+\mathcal{J}_2^2}}  & 0\\
                                  0 & 0 & 1 \\
                                \end{array}
                              \right)
\ee
which yield
\be
 \left(
                                \begin{array}{cccc}
                                \mathcal{J}'_1 \\
                                 \mathcal{J}'_2 \\
                                  \mathcal{J}'_3 \\
                                \end{array}
                              \right) = \
                             R_2 R_1 \left(
                                \begin{array}{cccc}
                                \mathcal{J}_1 \\
                                 \mathcal{J}_2 \\
                                  \mathcal{J}_3 \\
                                \end{array}
                              \right) =
                              \left(
                                \begin{array}{cccc}
                                0 \\
                                 0 \\
                                 \sqrt{\mathcal{J}_1^2+\mathcal{J}_2^2+\mathcal{J}_3^2} \\
                                \end{array}
                              \right).
\ee
This rotation acts on the orbital subspace $(\theta,\phi)$ via $\he'_r=R_2 R_1\he_r$, giving
\be \label{TPhi'}
\bal
&
\cos{\theta'}\=\he'_r\cdot\frac{\vec{\cJ}}{|\vec{\cJ}|}
\=\frac{\mathcal{J}_1 \sin{\theta}\cos{\phi}+\mathcal{J}_2 \sin{\theta}\sin{\phi}+\mathcal{J}_3\cos{\theta}}{\sqrt{\mathcal{J}_1^2+\mathcal{J}_2^2+\mathcal{J}_3^2}},
\\[4pt]
&
\tan{\phi'}\=\frac{(\mathcal{J}_1^2 + \mathcal{J}_2^2)\cot{\theta} - \mathcal{J}_3\,(\mathcal{J}_1 \cos{\phi} +\mathcal{J}_2 \sin{\phi})}{ (\mathcal{J}_1 \sin{\phi}- \mathcal{J}_2 \cos{\phi} ) \sqrt{
 \mathcal{J}_1^2+\mathcal{J}_2^2+\mathcal{J}_3^2}}.
\eal
\ee
Restricting the first formula to the mass shell, i.e.~making use of (\ref{charges}), one verifies that
\be \label{Tonshell}
\cos{\theta'(t)}\=-\frac{J_2}{\sqrt{\mathcal{J}_1^2+\mathcal{J}_2^2+\mathcal{J}_3^2}} \= -\frac{J_2}{\sqrt{2H}}\=\textrm{constant},
\ee
as it should be due to (\ref{cone}) above.

To figure out the time evolution of $\phi'$, one might insert the solutions (\ref{T}) and (\ref{Phi}) into the second line of (\ref{TPhi'}).
However, it is easier to employ the inverse transformation
\be
\bal
&
\cos{\theta}\=\frac{\mathcal{J}_3 \cos{\theta'}+\sqrt{\mathcal{J}_1^2 + \mathcal{J}_2^2}\,\sin{\theta'} \sin{\phi'}}{\sqrt{
 \mathcal{J}_1^2+\mathcal{J}_2^2+\mathcal{J}_3^2}},
\\[2pt]
&
\tan{\phi}\=\frac{\mathcal{J}_2 \sqrt{\mathcal{J}_1^2 + \mathcal{J}_2^2}\,\cot{\theta'} +
 (\mathcal{J}_1 \sqrt{\mathcal{J}_1^2+\mathcal{J}_2^2+\mathcal{J}_3^2}\,\cos{\phi'} -
    \mathcal{J}_2 \mathcal{J}_3 \sin{\phi'}) }{\mathcal{J}_1 \sqrt{\mathcal{J}_1^2 + \mathcal{J}_2^2}\,\cot{\theta'} -
 (\mathcal{J}_2 \sqrt{\mathcal{J}_1^2+\mathcal{J}_2^2+\mathcal{J}_3^2}\,\cos{\phi'} +
    \mathcal{J}_1 \mathcal{J}_3 \sin{\phi'}) } .
\eal
\ee
Using the first line and the result (\ref{Tonshell}) to express the evolution of $\sin\phi'(t)$ in terms of the solution (\ref{T}),
we readily see that
\be
\sin\phi'(t) \= \cos\bigl(\sqrt{2H}(t{-}t_0)\bigr) \qquad\Rightarrow\qquad
\phi'(t) \= \sqrt{2H}(t{-}t_1)
\ee
with a shifted integration constant $t_1$.
The result $\dot{\phi}^{\prime 2}=2H$ can also be inferred from $\dot{x}'_i\dot{x}'_i=\dot{x}_i\dot{x}_i$
using (\ref{Tonshell}) in the rotated frame and (\ref{uniformorbit}) in the unrotated one.

As compared to a free particle on the two-sphere, the presence of the spin degrees of freedom 
moves the circular orbit an azimuthal distance $-\frac{J_2}{\sqrt{2H}}$ away from the equatorial plane, 
while the angular velocity ${\dot\phi}'(t)=\sqrt{2H}$ is now linked to the energy of the {\it full\/} system.

The guiding principle to build transformation laws for the remaining variables $(p_\theta,p_\phi)$ and $(J_1,J_2,J_3)$ 
is to preserve the Poisson-structure relations (\ref{sr}). The corresponding partial differential equations seem to be intractable for the moment, 
indicating that more physical insight is needed. We plan to continue their study elsewhere.

\vspace{0.5cm}

\noindent
{\bf 6. Spinning extensions of the $\boldsymbol{D(2,1;\alpha)}$ superconformal mechanics}\\

\noindent
Any realization of $su(2)$ in Section~3 can be extended to a representation of the Lie superalgebra associated with $D(2,1;\alpha)$ \cite{AG}.
It is sufficient to introduce an extra bosonic canonical pair $(x,p)$ along with fermionic SU(2) spinor partners $(\psi_\alpha,\bar\p^\a)$ subject to ${(\psi_\alpha)}^{*}=\bar\p^\a$ for $\alpha=1,2$, and to extend (\ref{sr}) by the structure relations
\be\label{cr}
\{x,p\}=1 \und \{ \p_\a, \bar\p^\b \}=-\ic\,{\d_\a}^\b.
\ee
The generators of Lie superalgebra associated with $D(2,1;\alpha)$ read
\begin{align}\label{rep}
&
H=\frac{p^2}{2}+\frac{2 \alpha^2 }{x^2} \mathcal{J}_a \mathcal{J}_a+\frac{2 \alpha}{x^2} (\bar\p \s_a \p) \mathcal{J}_a -\frac{(1{+}2\alpha)}{4x^2} \p^2 \bar\p^2,
&& D=tH-\sfrac 12 x p,
\nonumber\\[4pt]
&
K=t^2 H-t x p +\sfrac 12 x^2,
&& \mathcal{I}_a=\mathcal{J}_a+\sfrac 12 (\bar\p \s_a \p),
\nonumber\\[2pt]
&
Q_\a=p \p_\a-\frac{2\ic\alpha}{x} {(\s_a \p)}_\a \mathcal{J}_a -\frac{\ic(1{+}2\alpha)}{2x} \bar\p_\a \p^2\ ,
&& S_\a=x \p_\a -t Q_\a,
\\[2pt]
&
\bar Q^\a =p \bar\p^\a+\frac{2\ic\alpha}{x} (\bar\p \s_a)^\a \mathcal{J}_a -\frac{\ic(1{+}2\alpha)}{2x} \p^\a \bar\p^2,
&& \bar S^\a=x \bar\p^\a -t \bar Q^\a,
\nonumber\\[4pt]
&
I_{-}=\sfrac{\ic}{2} \psi^2, \qquad \qquad \qquad \qquad I_{+}=-\sfrac{\ic}{2} {\bar\psi}^2,
&& I_3=\sfrac12 \bar\psi \psi,
\nonumber
\end{align}
where $\s_a$ are the Pauli matrices. When verifying the structure relations of the superalgebra (see the Appendix), one only needs to use the bracket $\{\mathcal{J}_i,\mathcal{J}_j\}=\epsilon_{ijk} \mathcal{J}_k$ and the fact that the $\mathcal{J}_i$ commute with $(x,p,\psi_\alpha,\bar\p^\a)$ without specifying the actual content of $\mathcal{J}_i$. As far as dynamical realizations are concerned, $H$ is interpreted as the Hamiltonian. $D$ and $K$ are treated as the generators of dilatations and special conformal transformations. $Q_\a$ are the supersymmetry generators and $S_\a$ are their superconformal partners. $\mathcal{I}_a$ generate the $R$-symmetry subalgebra $su(2)$. So do also $I_{\pm}$ and $I_3$ for which the Cartan basis is chosen.

A few comments are in order. An attempt to accommodate spin degrees of freedom within $D(2,1;\alpha)$ superconformal mechanics was made in \cite{FIL} (see also related earlier work \cite{FIL1}). Bosonic SU(2) doublet variables $(z_\alpha, {\bar z}^\alpha)$ with ${(z_\alpha)}^{*}={\bar z}^\a$ and $\alpha=1,2$ have been introduced, which obey the bracket $\{ z_\alpha, {\bar z}^\beta \}=-\ic{\delta_\alpha}^\beta$ and give rise to the $su(2)$ generators
\be\label{nd}
J^{\alpha\beta}\=\sfrac{\ic}{2} (z^\alpha {\bar z}^\beta+z^\beta {\bar z}^\alpha ).
\ee
As the extra variables parametrize a two-dimensional sphere,
\be
z_1=r \cos\sfrac{\theta}{2}\,\eu^{\ic\frac{\phi}{2}}, \qquad z_2=r \sin\sfrac{\theta}{2}\,\eu^{-\ic\frac{\phi}{2}}, \qquad z_\alpha {\bar z},^\alpha=r^2
\ee
and their dynamics is governed by the Wess--Zumino-type action
\be
S\=-\frac{r^2}{2}\int\!\diff t \,\dot\phi\,\cos{\theta} ,
\ee
one concludes that $(z_\alpha, {\bar z}^\alpha)$ are non-propagating harmonic variables \cite{FIL}. This is to be contrasted with the fully fledged spin dynamics resulting form the $su(2)$ realizations in Section~3.

A generalization of \cite{FIL,FIL1} to the many-body case was proposed in \cite{KL}. According to the analysis in \cite{AG}, the algebraic construction in \cite{KL} remains valid if one replaces $(\ref{nd})$ by any other dynamical realization of $su(2)$, provided the kinetic term entering the resulting Hamiltonian involves a non-degenerate metric. Combining the results in \cite{AG} with those in Section~3 one can readily build a spinning extension of $D(2,1;\alpha)$ superconformal many-body mechanics based upon any chosen solution of the generalized Witten--Dijkgraaf--Verlinde--Verlinde equations \cite{KL}, including the $\vee$-system solutions proposed recently in \cite{AF}.

Finally, by introducing an extra fermionic canonical pair $(\chi_\alpha,\bar\chi^\a)$ with
$\bar\chi^\a={(\chi_\alpha)}^{*}$ for $\alpha=1,2$, and by incorporating the bracket
\be
\{ \chi_\a, \bar\chi^\b \}=-\ic\,{\d_\a}^\b
\ee
into the Poisson structure,
one can further generalize the $su(2)$ generators of Section~3 via
\be\label{s4}
\mathcal{J}_a \quad \rightarrow \quad \mathcal{J}_a+\sfrac12 (\bar\chi \sigma_a \chi).
\ee
The resulting model (\ref{rep}) will describe the coupling of spinning $D(2,1;\alpha)$ superconformal mechanics
to an extra on-shell  type-$(0,4,4)$ supermultiplet realized in terms of $\chi_\alpha$ and $\bar\chi^\a$.

\vspace{0.5cm}

\noindent
{\bf 7. Conclusions}\\

\noindent
To summarize, in this work we have built spinning extensions the $D(2,1;\alpha)$ superconformal mechanics by properly adjusting
the SU(2) generators associated with the model of a relativistic spinning particle coupled to spherically symmetric four-dimensional Einstein--Maxwell backgrounds. The spin degrees of freedom are represented by a symmetric Euler top.
A peculiar feature of the construction is the non-standard Poisson structure inherited from the parent relativistic formulation \cite{AKH}. It was shown that the compact sector of the spinning
$D(2,1;\alpha)$ superconformal mechanics describes either a particle moving on a two-dimensional sphere coupled to a symmetric Euler top, or the same system in the external field of a Dirac monopole, or a particle propagating on the group manifold of SU(2) interacting with a symmetric Euler top. Each case was proven to be superintegrable, and the general solution to the equations of motion was constructed. A possible generalization of the analysis to the many-body case has been discussed.

There are several directions in which the present work can be continued. The fermionic degrees of freedom $(\psi_\alpha,\bar\p^\a)$ introduced in Section~6 represent supersymmetric partners for the bosonic variables $(x,\theta,\phi,\xi)$. Together they form on-shell supermultiplets of the $D(2,1;\alpha)$ supergroup. It will be interesting to study whether the fermionic counterparts can also be associated with spin degrees of freedom in the spirit of recent work \cite{KLS}.

One can consider a relativistic spinning particle on more general backgrounds and in an arbitrary dimension. The associated reduced angular sector might be of interest with regard to its integrability. The Myers--Perry black-hole geometry with its SU($n$) isometry is a case to study.

A systematic investigation of the angular part of generic many-body conformal mechanics was initiated in \cite{HKLN,HLNS}. The construction of spinning extensions of a generic spherical mechanical system is an intriguing open problem.

\vspace{0.5cm}

\noindent{\bf Acknowledgements}\\

\noindent
A.G. is grateful to the Institut f\"ur Theoretische Physik at Leibniz Universit\"at Hannover for the hospitality extended to him at the initial stage of this project. This work was supported by the Tomsk Polytechnic University competitiveness enhancement program.

\vspace{0.5cm}

\noindent
{\bf Appendix}

\vspace{0.5cm}
\noindent
The standard orthonormal basis for spherical coordinates have the Cartesian components
\be
\begin{pmatrix} \he_r \end{pmatrix} =
\begin{pmatrix} x_1 \\ x_2 \\ x_3 \end{pmatrix}  =
\begin{pmatrix} \sin\theta\cos\phi \\ \sin\theta\sin\phi \\ \cos\theta \end{pmatrix} \ ,\qquad
\begin{pmatrix} \he_\theta \end{pmatrix} =
\begin{pmatrix} \cos\theta\cos\phi \\ \cos\theta\sin\phi \\ -\sin\theta \end{pmatrix} \ ,\qquad
\begin{pmatrix} \he_\phi \end{pmatrix} =
\begin{pmatrix} -\sin\phi \\ \cos\phi \\ 0 \end{pmatrix} \ .
\nonumber
\ee

The structure relations of the Lie superalgebra associated with the exceptional supergroup $D(2,1;\alpha)$ read
\begin{align}
&
\{ H,D \}=H, && \{ H,K \}=2D,
\nonumber\\[2pt]
&
\{D,K\}=K, && \{ \mathcal{I}_a,\mathcal{I}_b \}=\epsilon_{abc} \mathcal{I}_c,
\nonumber\\[2pt]
&
\{ Q_\a, \bar Q^\b \}=-2 \ic H {\d_\a}^\b, &&
\{ Q_\a, \bar S^\b \}=-2\alpha (\s_a)_\a^{\ \b} \mathcal{I}_a+2\ic D {\d_\a}^\b+2(1{+}\alpha)I_3 {\d_\a}^\b,
\nonumber\\[2pt]
&
\{ S_\a, \bar S^\b \}=-2\ic K {\d_\a}^\b, &&
\{ \bar Q^\a, S_\b \}=2\alpha(\s_a)_\b^{\ \a} \mathcal{I}_a+2\ic D {\d_\b}^\a-2(1{+}\alpha)I_3 {\d_\b}^\a,
\nonumber\\[2pt]
&
\{ Q_\a, S_\b \}=2\ic (1{+}\alpha) \epsilon_{\alpha \beta} I_{-}, &&
\{ {\bar Q}^\a, {\bar S}^\b \}=-2\ic (1{+}\alpha) \epsilon^{\alpha \beta} I_{+},
\nonumber\\[2pt]
&
\{ D,Q_\a\} = -\sfrac{1}{2} Q_\a, && \{ D,S_\a\} =\sfrac{1}{2} S_\a,
\nonumber\\[2pt]
&
\{ K,Q_\a \} =S_\a, && \{ H,S_\a \}=-Q_\a,
\nonumber\\[2pt]
&
\{ \mathcal{I}_a,Q_\a\} =\sfrac{\ic}{2} (\s_a)_\a^{\ \b} Q_\b, && \{ \mathcal{I}_a,S_\a\} =\sfrac{\ic}{2} (\s_a)_\a^{\ \b} S_\b,
\nonumber\\[2pt]
& \{ D,\bar Q^\a \} =-\sfrac{1}{2} \bar Q^\a, && \{ D,\bar S^\a\} =\sfrac{1}{2} \bar S^\a,
\nonumber\\[2pt]
& \{K,\bar Q^\a\} =\bar S^\a, && \{ H,\bar S^\a\} =-\bar Q^\a,
\nonumber
\end{align}
\begin{align}
&
\{\mathcal{I}_a,\bar Q^\a\} =-\sfrac{\ic}{2} \bar Q^\b (\s_a)_\b^{\ \a}, && \{ \mathcal{I}_a,\bar S^\a\} =-\sfrac{\ic}{2}
\bar S^\b (\s_a)_\b^{\ \a},
\nonumber\\[2pt]
&
\{ I_{-},\bar Q^\a \} =\epsilon^{\alpha \beta} Q_\beta, && \{ I_{-},\bar S^\a \} =\epsilon^{\alpha \beta} S_\beta,
\nonumber\\[2pt]
&
\{ I_{+},Q_\a\} =-\epsilon_{\alpha\beta} \bar Q^\b, && \{ I_{+},S_\a\} =-\epsilon_{\alpha\beta} \bar S^\b,
\nonumber\\[2pt]
&
\{ I_3,Q_\a \} =\sfrac{\ic}{2} Q_\a, &&  \{ I_3,S_\a \} =\sfrac{\ic}{2} S_\a,
\nonumber\\[2pt]
&
\{ I_3,\bar Q^\a\} =-\sfrac{\ic}{2} \bar Q^\a, &&  \{ I_3,\bar S^\a \} =-\sfrac{\ic}{2} \bar S^\a,
\nonumber\\[2pt]
&
\{ I_{-},I_3\} =-\ic I_{-}, &&  \{ I_{+},I_3 \} =\ic I_{+},
\nonumber\\[2pt]
&
\{ I_{-},I_{+}\} =2\ic I_{3}. &&
\nonumber
\end{align}
The Pauli matrices $\bigl((\s_a)_\a^{\ \b}\bigr)$
are chosen in the form
\be
\s_1=\begin{pmatrix}0 & 1\\
1 & 0
\end{pmatrix}\ , \qquad \s_2=\begin{pmatrix}0 & -\ic\\
\ic & 0
\end{pmatrix}\ ,\qquad
\s_3=\begin{pmatrix}1 & 0\\
0 & -1
\end{pmatrix}\ ,
\nonumber
\ee
which obey
\bea
&&
(\s_a \s_b)_\a^{\ \b} +(\s_b \s_a)_\a^{\ \b}=2 \d_{ab} {\d_\a}^\b \ , \quad
(\s_a \s_b)_\a^{\ \b} -(\s_b \s_a)_\a^{\ \b}=2\ic \e_{abc} (\s_c)_\a^{\ \b} \ ,
\nonumber\\[2pt]
&&
(\s_a \s_b)_\a^{\ \b}=\d_{ab} {\d_\a}^\b +\ic \e_{abc} (\s_c)_\a^{\ \b} \ , \quad
(\s_a)_\a^{\ \b} (\s_a)_\g^{\ \r}=2 {\d_\a}^\r {\d_\g}^\b-{\d_\a}^\b {\d_\g}^\r\ ,
\nonumber\\[2pt]
&&
(\s_a)_\a^{\ \b} \e_{\b\g} =(\s_a)_\g^{\ \b} \e_{\b\a}\ , \quad
\e^{\a\b} (\s_a)_\b^{\ \g}=\e^{\g\b} (\s_a)_\b^{\ \a} \ ,
\nonumber
\eea
In Section~6 lower Greek indices designate SU(2) doublet representations. Complex conjugation
yields equivalent representations to which one assigns upper indices, ${(\p_\a)}^{*}={\bar\p}^\a$.
Spinor indices are raised and lowered with the use of the SU(2)-invariant
antisymmetric matrices $\epsilon$,
\be
\p^\a=\e^{\a\b}\p_\b\ , \quad {\bar\p}_\a=\e_{\a\b} {\bar\p}^\b\ ,
\nonumber
\ee
where $\e_{12}=1$ and $\e^{12}=-1$. For spinor bilinears we stick to the notation
\be
\quad \p^2=(\p^\a \p_\a)\ , \quad
\bar\p^2=(\bar\p_\a \bar\p^\a )\ , \quad \bar\p \p=(\bar\p^\a \p_\a )\ ,
\nonumber
\ee
such that
\be
\p_\a \p_\b=\sfrac 12 \e_{\a\b} \p^2\ , \quad \bar\p^\a \bar\p^\b=\sfrac 12 \e^{\a\b} \bar\p^2\ , \quad
\p_\a \bar\p_\b-\p_\b \bar\p_\a=\e_{\a\b}\ , \quad \bar\p\,\s_a \p=\bar\p^\a (\s_a)_\a^{\ \b} \p_\b\ .
\nonumber
\ee
Our conventions for complex conjugation read
\bea
&&
{(\psi_\alpha)}^{*}=\bar\psi^\alpha\ , \qquad {(\bar\psi_\alpha)}^{*}=-\psi^\alpha\ , \qquad
{(\psi^2)}^{*}=\bar\psi^2\ , \qquad {(\bar\psi\,\sigma_a \chi)}^{*}=\bar\chi \sigma_a \psi\ .
\nonumber
\eea

\end{document}